\documentclass[aps,prb,twocolumn,amssymb,amsmath,showpacs,superscriptaddress]{revtex4-1}
\usepackage{bm}
\usepackage{times}
\usepackage{graphicx}
\usepackage{color}
\usepackage{dcolumn}
\usepackage[colorlinks=true, pdfstartview=FitV, linkcolor=blue, citecolor=blue, urlcolor=blue]{hyperref}
\usepackage{appendix}
\usepackage[normalem]{ulem}

\begin{document}

\title{Linear thermal noise induced by Berry curvature dipole in a four-terminal system}

\author{Wenyu Chen}
\thanks{These authors contributed equally to this work.}
\affiliation{College of Physics and Optoelectronic Engineering, Shenzhen University, Shenzhen 518060, China}

\author{Miaomiao Wei}
\thanks{These authors contributed equally to this work.}
\affiliation{College of Physics and Optoelectronic Engineering, Shenzhen University, Shenzhen 518060, China}


\author{Yunjin Yu}
\affiliation{College of Physics and Optoelectronic Engineering, Shenzhen University, Shenzhen 518060, China}

\author{Fuming Xu}
\email[]{xufuming@szu.edu.cn}
\affiliation{College of Physics and Optoelectronic Engineering, Shenzhen University, Shenzhen 518060, China}
\affiliation{Quantum Science Center of Guangdong-Hongkong-Macao Greater Bay Area (Guangdong), Shenzhen 518045, China}

\author{Jian Wang}
\email[]{jianwang@hku.hk}
\affiliation{College of Physics and Optoelectronic Engineering, Shenzhen University, Shenzhen 518060, China}
\affiliation{Quantum Science Center of Guangdong-Hongkong-Macao Greater Bay Area (Guangdong), Shenzhen 518045, China}
\affiliation{Department of Physics, The University of Hong Kong, Pokfulam Road, Hong Kong, China}

\begin{abstract}

In this work, we numerically investigate linear thermal noise in a four-terminal system with a finite Berry curvature dipole (BCD) using the nonequilibrium Green's function formalism. By comparing with the semiclassical results for bulk systems, we establish a one-to-one correspondence between terminal-resolved linear noise in multi-terminal systems and direction-resolved noise in bulk transport. Specifically, the auto-correlation function scales as $2 k_B T$ when the driving field is perpendicular to the BCD and vanishes when they are parallel, whereas the cross-correlation scales as $k_B T$. Both the auto- and cross-correlation functions exhibit pronounced peaks near the band edges, consistent with BCD-induced features. In addition, the linear thermal noise increases approximately linearly with $T$ at low temperatures and is suppressed by dephasing effect at high temperatures. Our work bridges semiclassical bulk theory and quantum multi-terminal theory for linear thermal noise, highlighting the symmetry(geometry)-selection rule in quantum transport.

\end{abstract}
\maketitle

\section{Introduction}

Quantum geometric quantities, including quantum metric and Berry curvature, provide a compact and physically transparent framework that connects the wave-function geometry to measurable optical and electronic responses of quantum materials.~\cite{Xiao2010,Torma2023PRLQG,LiuLu2025NSR_QG,Kang2025QGT,Yu2025QGReview} Berry curvature underpins Hall-type charge transport, such as the intrinsic anomalous Hall effect and various nonlinear Hall effects.~\cite{Nagaosa2010,DuLuXie2021NRP} Quantum metric governs phenomena involving interband coherence, such as orbital magnetic susceptibility and nonlinear optical processes.~\cite{Piechon2016,MorimotoNagaosa2016,Ahn2020} These effects induced by quantum geometry are often constrained by symmetry, thereby leading to symmetry-selection rules for the allowed responses. For example, the intrinsic anomalous Hall effect induced by Berry curvature requires broken time-reversal ($\cal T$) symmetry,~\cite{Nagaosa2010} whereas Berry curvature dipole can induce the extrinsic second-order nonlinear Hall effect in $\cal T$-invariant but inversion-broken systems.~\cite{SodemannFu2015,Guinea1}

Recently, the language of quantum geometry has been extended to describe quantum noise, including both thermal noise and shot noise,~\cite{Blanter2000} of charge current and photocurrent.~\cite{Neupert2013} Geometric quantities that govern charge and optical currents can also influence their noise spectra, reflecting the symmetry constraints of the underlying system. For example, Berry curvature dipole (BCD) can induce an intrinsic linear thermal noise in $\cal T$-invariant systems, in dual relation with the BCD-induced second-order Hall effect.~\cite{Wei2023} In the optical regime, quantum geometry can generate distinct features in the shot noise of shift and injection current.~\cite{NagaosaSN1} Meanwhile, shot noise of photocurrents can serve as a sensitive probe of geometric characteristics in both nonmagnetic ($\cal T$-invariant) and magnetic ($\cal T$-broken) quantum materials.~\cite{xiangDSN,L-Zhang} Moreover, in $\cal PT$ symmetric systems with vanishing Berry curvature, quantum metric can directly contribute to nonlinear thermal noise.~\cite{Agarwal25} Despite these advances, a comprehensive understanding of how quantum geometry affects the transport behaviors of quantum noise in multi-terminal systems remains lacking.

Conventional quantum transport theories, such as scattering matrix theory and the nonequilibrium Green's function formalism (NEGF),~\cite{Datta1995,HaugJauho2008} are built on the Landauer-B\"{u}ttiker framework,~\cite{Landauer1957,Landauer1970,Buttiker1986,MeirWingreen1992} where contributions from quantum geometry to transport properties are not transparent.~\cite{Xu2025} Therefore, incorporating quantum geometry into quantum multi-terminal transport theory is both challenging and timely. In this work, we develop a quantum multi-terminal transport theory for linear thermal noise based on NEGF, which satisfies both current conservation and gauge invariance. Then we apply this theory to numerically investigate linear thermal noise in a four-terminal system and elucidate the role of the BCD. By comparing with the semiclassical results for bulk transport,~\cite{Wei2023} we establish a one-to-one correspondence between terminal-resolved noise in multi-terminal systems and direction-resolved noise in bulk transport. This correspondence reveals symmetry-imposed selection rules for linear thermal noise: the auto-correlation is finite when the driving field is perpendicular to the BCD and vanishes when they are parallel, while the cross-correlation exhibits a distinct scaling behavior. We further analyze the dependence of linear thermal noise on Fermi energy and temperature, and identify BCD-related enhancement near the band edges. We also demonstrate that the dephasing effect suppresses noise at high temperatures.

The rest of this paper is organized as follows. Section.~II introduces the model Hamiltonian and outlines the quantum theory for calculating linear thermal noise in multi-terminal systems. Section.~III presents the numerical results and comparison with semiclassical predictions for bulk systems. Finally, Section.~IV summarizes our conclusions.

\section{Model and Formalism}

\subsection{Model Hamiltonian}

\begin{figure}[tbp]
\centering
\includegraphics[width=\columnwidth]{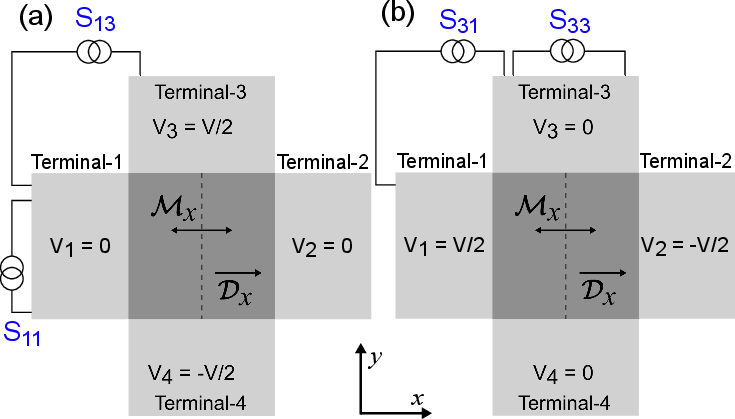}
\caption{Schematic of a two-dimensional(2D) four-terminal Hall setup for measuring quantum noises. The voltage profile ${\cal{V}}=(V_1,V_2,V_3,V_4)$ describes the bias voltages applied to each terminal. (a) In setup {\color{blue}I}, ${\cal{V}}=(0,0,V/2,-V/2)$, corresponding to a driving electric field along the $y$ direction. (b) In setup {\color{blue}II}, ${\cal{V}}=(V/2,-V/2,0,0)$, corresponding to a driving field along the $x$ direction. The four-terminal system is modeled by the tight-binding Hamiltonian in Eq.~(\ref{Ham2}). Here $\mathcal{M}_x$ denotes the mirror symmetry in the $x$ direction, and $\mathcal{D}_x$ is the pseudovector representing the Berry curvature dipole. $S_{11}$ and $S_{33}$ are the auto-correlation function of the current in Terminal-1 and Terminal-3, respectively. $S_{13}$ and $S_{31}$ are the cross-correlation between currents in Terminal-1 and Terminal-3.}\label{fig1}
\end{figure}

The typical Hamiltonian for studying Berry curvature dipole (BCD) related physics is the following two-dimensional(2D) tilted massive Dirac Hamiltonian,~\cite{Papaj2019,Debttam2020,Pankaj2022}
\begin{eqnarray}
H(\textbf{k}) = A k^2 + (Bk^2 + \delta)\sigma_z + v_y k_y\sigma_y + D\sigma_x, \label{Ham1}
\end{eqnarray}
where $A$, $B$, $\delta$, $v_y$, and $D$ are material-dependent parameters. $k^2 = k_x^2 + k_y^2$ and $\sigma_{x/y/z}$ are Pauli matrices. The tilted term $v_y k_y\sigma_y$ breaks mirror symmetry ${\cal{M}}_y$ and hence inversion symmetry of this Hamiltonian, i.e., $H(k_x,k_y)\neq H(k_x,-k_y)$. The system preserves $\cal{T}$ symmetry and mirror symmetry ${\cal{M}}_x$ ($H(k_x,k_y)= H(-k_x,k_y)$). As a result, this Hamiltonian possesses a BCD pseudovector along the $x$ direction, labeled as ${\cal{D}}_x$ in Fig.~\ref{fig1}.

In tight-binding representation, the Hamiltonian in Eq.~(\ref{Ham1}) can be projected on a square lattice in the $x$-$y$ plane as~\cite{Wei2022}
\begin{eqnarray}
H=\sum_{\textbf{i}}[\psi_\textbf{i}^\dagger T_0 \psi_\textbf{i} + \psi_\textbf{i}^\dagger T_x \psi_{\textbf{i}+\textbf{a}_x} + \psi_\textbf{i}^\dagger T_y \psi_{\textbf{i}+\textbf{a}_y} +\text{h.c.}], \label{Ham2}
\end{eqnarray}
with
\begin{align*}
     & T_0=-4T_x+\delta\sigma_z+D\sigma_x, \\
     & T_x=-(AI_0+B\sigma_z)/a_0^2,          \\
     & T_y=T_x-iv_y\sigma_y/(2a_0).
\end{align*}
Here $\psi_\textbf{i}^\dagger$ is the creation operator at site $\textbf{i}$ with $\textbf{i}=(\textbf{i}_x,\textbf{i}_y)$ indexing the lattice site. $\textbf{a}_x=(a_0,0)$ and $\textbf{a}_y=(0,a_0)$ are the unit vectors in the $x$ and $y$ directions, respectively, with $a_0$ the lattice spacing between adjacent sites. $I_0$ is a $2 \times 2$ identity matrix. In the calculation, the parameters are chosen as $A=0$, $B=1$, $\delta=-0.25$, $v_y=1$, $D=0.1$, and $a_0=1$.~\cite{Debttam2020,Pankaj2022,Wei2022} The energy unit is $eV$.

\subsection{Linear thermal noise induced by Berry curvature dipole in bulk systems}

In terms of the current density operator ${\hat{J}}_a$, quantum noise of a bulk system is defined as~\cite{Wei2023}
\begin{eqnarray}
{\cal{S}}_{ab} = \langle \Delta {\hat{J}}_a \Delta {\hat{J}}_b \rangle,
\end{eqnarray}
where $\Delta {\hat{J}}_a = {\hat{J}}_a - \langle {\hat{J}}_a \rangle$ denotes the fluctuation of the current density. ${a, b} \in \{x, y, z \}$ label the spatial directions.


For the Hamiltonian in Eq.~(\ref{Ham1}) with a finite Berry curvature dipole ${\cal D}_x$, Ref.~[\onlinecite{Wei2023}] derived the linear thermal noises of this bulk system as~\cite{Wei2023}
\begin{eqnarray}
{\cal S}^{(1)}_{xx} &=& 2k_B T {\cal D}_x E_y, \label{Sxx} \\
{\cal S}^{(1)}_{yy} &=& 0, \label{Syy} \\
{\cal S}^{(1)}_{xy} &=& -k_B T {\cal D}_x E_x. \label{Sxy}
\end{eqnarray}
Here $E_x$($E_y$) denotes the electric field applied along the $x$($y$) direction, $k_B$ is the Boltzmann constant, and $T$ is the temperature. ${\cal S}^{(1)}_{xx}$(${\cal S}^{(1)}_{yy}$) is the auto-correlation function of the current density in $x$($y$) direction, while ${\cal S}^{(1)}_{xy}$ is the cross-correlation function between $x$ and $y$ components. In contrast to the BCD-induced extrinsic second-order Hall effect,~\cite{SodemannFu2015} ${\cal S}^{(1)}_{xx}$ in response to $E_y$ is the intrinsic linear Hall signal in $\cal T$-invariant systems,~\cite{Wei2023} which is the intrinsic manifestation of BCD.~\cite{Xu2025}

Equation~(\ref{Sxy}) shows that the linear thermal noise is directly connected to the BCD-induced second-order nonlinear Hall current $J_H^{(2)} = - \tau {\cal D}_x E_x^2$,~\cite{SodemannFu2015,Wei2023} which shows
\begin{equation}
\frac{\partial {\cal S}^{(1)}_{xy}}{\partial E_x} = - k_B T \frac{\partial^2 J_H^{(2)}}{\partial E_x^2}. \label{FD1}
\end{equation}
Here $J_H^{(2)}$ is the second-order current density and $\tau$ is the relaxation time. Recall that in quantum transport through a two-terminal system, the fluctuation--dissipation theorem gives a general relation~\cite{Tobiska,Kamenev}
\begin{equation}
\frac{\partial S^{(1)}}{\partial V} = - k_B T \frac{\partial^2 I}{\partial V^2}, \label{FD2}
\end{equation}
where $S^{(1)}$ is the linear thermal noise defined as the correlation of terminal currents (see Eq.~(\ref{S00}) in the next subsection), and $\partial^2 I / \partial V^2$ is the second derivative of current $I$ with respect to the bias voltage $V$.~\cite{Tobiska,Kamenev} The similarity between Eq.~(\ref{FD1}) and Eq.~(\ref{FD2}) reveals a close correspondence between the linear thermal noises in bulk systems, ${\cal S}^{(1)}_{ab}$ defined in terms of current density, and those in multi-terminal systems, $S_{\alpha \beta}^{(1)}$ defined with terminal currents. We will further explore the correspondence between them in the following sections.

\subsection{Linear thermal noise in multi-terminal systems}

For a multi-terminal system, such as the Hall setup in Fig.~\ref{fig1}, quantum noise of charge currents is defined as the correlation of current fluctuation between terminals,~\cite{Blanter2000,Buttiker1992,SukhorukovLoss1999PRB}
\begin{equation}
S_{\alpha\beta}\equiv \frac{1}{2}\langle \Delta\hat{I}_\alpha \Delta\hat{I}_\beta + \Delta\hat{I}_\beta \Delta\hat{I}_\alpha\rangle, \label{S00}
\end{equation}
where $\langle\dots\rangle$ denotes statistical average and $\Delta \hat{I}_\alpha=\hat{I}_\alpha-\langle\hat{I}_\alpha\rangle$ is the current fluctuation at terminal $\alpha$. After Fourier transformation, the zero-frequency quantum noise is expressed in terms of the scattering matrix as ($\hbar= e = 1$),~\cite{Blanter2000}
\begin{equation}
\begin{aligned}
S_{\alpha\beta}(0) & =\sum_{\gamma\delta} \int_{E} \text{Tr}[A_{\gamma\delta}(\alpha,E) A_{\delta\gamma}(\beta,E)]  \\
                                                 & \times[f_\gamma(1-f_\delta)+f_\delta(1-f_\gamma)], \label{Snoise0}
\end{aligned}
\end{equation}
where $A_{\gamma\delta}(\alpha,E) = \delta_{\gamma \alpha} \delta_{\delta\alpha} - s^\dagger_{\alpha\gamma}(E) s_{\alpha\delta}(E)$, and $\alpha, \beta, \gamma, \delta$ label the terminals. The scattering matrix element $s_{\alpha\beta}$ connects incoming modes from terminal $\beta$ to outgoing modes in terminal $\alpha$. $f_\beta = f_\beta(E + V_\beta)$ is the Fermi distribution function of terminal $\beta$ with a bias voltage $V_\beta$, which reduces to the equilibrium distribution $f_0$ at zeros bias.

Since thermal noise is dominant in the low voltage regime $e V \ll k_B T$, we expand $S_{\alpha\beta}$ with respect to the bias voltage and obtain
\begin{equation}
    S_{\alpha\beta}=S_{\alpha\beta}^{(0)} + S_{\alpha\beta}^{(1)}(V) + \mathcal{O}(V^2) \dots \label{Snoise1}
\end{equation}
Here $S_{\alpha\beta}^{(0)}$ is the equilibrium Nyquist-Johnson noise,~\cite{Buttiker1990PRLNoise} arising from thermal fluctuations of occupation numbers in the electron reservoirs.~\cite{Blanter2000} $S_{\alpha\beta}^{(1)}$ is the linear thermal noise, which is the main focus of this work.

To satisfy gauge invariance, it is essential to include the internal Coulomb response of the electronic system to the external bias.~\cite{Buttiker1992,Buttiker1993,Wei2022,Wei-3H} For this purpose, we employ the Fisher--Lee relation~\cite{Fisher1981} $s_{\alpha\beta}=-\delta_{\alpha\beta}+i\Gamma^{1/2}_{\alpha}G^r\Gamma^{1/2}_\beta$ and obtain
\begin{equation}
s^\dagger_{\alpha\beta}s_{\alpha\beta}=\delta_{\alpha\beta}+\Gamma_\alpha G^r\Gamma_\beta G^a-\delta_{\alpha\beta}\Gamma_\alpha G^r\Gamma G^a, \nonumber
\end{equation}
where the internal Coulomb potential $U$ enters explicitly through the retarded Green's function,
\begin{equation}
G^r=\frac{1}{E - H - U -\Sigma^r}. \label{Green}
\end{equation}
Here $\Sigma^r = \sum_\alpha \Sigma^r_\alpha$ is the self-energies from the terminals. $\Gamma = \sum_\alpha \Gamma_\alpha$ and $\Gamma_\alpha = i[\Sigma^r_\alpha - (\Sigma^r_\alpha)^\dagger]$ is the linewidth function, which describes the coupling between terminal $\alpha$ and the central region.

To obtain the linear thermal noise, it is enough to expand $A_{\gamma\delta}(\alpha)$ in Eq.~(\ref{Snoise0}) and hence the Green's function in Eq.~(\ref{Green}) to linear order in the bias voltage $V$. From the Dyson equation and under the wideband limit, we have
\begin{eqnarray}
G^r = G^r_0 + G^r_0 U G^r_0 + \cdots \nonumber
\end{eqnarray}
where $G^r_0 = (E-H-\Sigma^r)^{-1}$ is the equilibrium Green's function.
The internal Coulomb potential is expressed as $U=\sum_{\alpha}u_{\alpha}V_\alpha$,~\cite{Wei2022} where $u_\alpha$ is the characteristic potential describing the first-order internal response to the external bias. $u_\alpha$ is defined below in Eq.~(\ref{u1}). Physically, when bias-driven electrons are injected into the system, a nonequilibrium charge distribution is formed due to the long-range Coulomb interaction, which induces the internal Coulomb potential. This induced Coulomb potential maintains gauge invariance for nonlinear transport.~\cite{but22,Wei2022,Wei-3H,ma1,wbg2} The Fermi distribution is also expanded to linear order in $V$,~\cite{Wei2023}
\begin{eqnarray}
f_\gamma(1-f_\delta)+f_\delta(1-f_\gamma) = 2g_2 - g_2' (V_\gamma +V_\delta), \label{fexpand}
\end{eqnarray}
where $g_2= f_0(1-f_0)$ and $g_2'= (1-2f_0) f_0'$ with $f_0'=\partial_E f_0$.

Substituting these expansions into Eq.~(\ref{Snoise0}), the linear thermal noise is expressed as
\begin{eqnarray}
S^{(1)}_{\alpha\beta} & = 2\sum_{\gamma} \int_{E} {\rm{Tr}} \{ {[ {{\partial_E}( - {{\delta _{\alpha \gamma }}{{\tilde G}_{\beta \alpha }} - {\delta _{\beta \gamma }}{{\tilde G}_{\alpha \beta}} + {\delta_{\alpha \beta}}{{\tilde G}_{\alpha \gamma}}} )} ]}\nonumber \\
& +{[ {{\partial_{V_\gamma}}( - {{\delta _{\alpha \gamma}}{{\tilde G}_{\beta \alpha}} - {\delta_{\beta \gamma}}{{\tilde G}_{\alpha \beta}} + {\delta_{\alpha \beta}}{{\tilde G}_{\alpha \gamma}}} )} ]} \}  g_2 V_\gamma, \label{Snoise4}
\end{eqnarray}
where $\tilde{G}_{\alpha\beta}= \Gamma_\alpha G_0^r (\Gamma_\beta -\delta_{\alpha\beta} \Gamma) G_0^a$ and $\partial_{V_\gamma} \tilde{G}_{\alpha\beta} = \Gamma_\alpha G_0^r [ u_\gamma G_0^r(\Gamma_\beta -\delta_{\alpha\beta} \Gamma ) + (\Gamma_\beta -\delta_{\alpha\beta} \Gamma)G_0^a u_\gamma ] G_0^a$. $G_0^a = (G_0^r)^\dagger$ is the advanced Green's function. Throughout the derivation, we have adopted the wideband approximation. In this expression, the first term represents the external contribution from the injected electrons, and the second term is the internal contribution from the induced Coulomb potential. One can verify that $\sum_{\beta}S^{(1)}_{\alpha \beta} = \sum_{\alpha} S^{(1)}_{\alpha \beta} = 0$, which ensures current conservation.

The characteristic potential $u_\alpha$ is essential to preserve the gauge invariance of $S^{(1)}_{\alpha \beta}$. In general, $u_\alpha$ is obtained by solving the Poisson equation~\cite{Buttiker1993,wbg2,but22,zhanglei}
\begin{equation}
-\nabla^2 u_\alpha = 4\pi q^2 \frac{dn_\alpha}{dE} - 4\pi q^2 \frac{dn}{dE} u_\alpha, \label{u1}
\end{equation}
which usually requires self-consistent iteration. A widely used simplification is the quasi-neutrality approximation~\cite{Buttiker1993,Wei2022,YDWei2009,zhanglei,NJP2023}, which assumes that local charge density is zero at each lattice site. Under this approximation, Eq.~(\ref{u1}) reduces to $-\nabla^2 u_\alpha = 0$, which leads to
\begin{eqnarray}
u_\alpha=\frac{dn_\alpha}{dE}/\frac{dn}{dE}.
\end{eqnarray}
Here $dn_\alpha/dE$ is the injectivity from terminal $\alpha$,~\cite{Buttiker1993,but3}
\begin{eqnarray}
\frac{dn_{\alpha}(x)}{dE} = -\int_E \partial_E f [G^r_0 \Gamma_{\alpha} G^a_0]_{xx}, \label{inj}
\end{eqnarray}
and the summation of injectivities over all terminals gives the local density of states,~\cite{but22,wbg2,Buttiker1993}
\begin{eqnarray}
\sum_\alpha \frac{dn_{\alpha}}{dE} = \frac{dn}{dE}. \label{ldos}
\end{eqnarray}


\section{NUMERICAL RESULTS AND DISCUSSION}

\begin{figure}[tbp]
\centering
\includegraphics[width=\columnwidth]{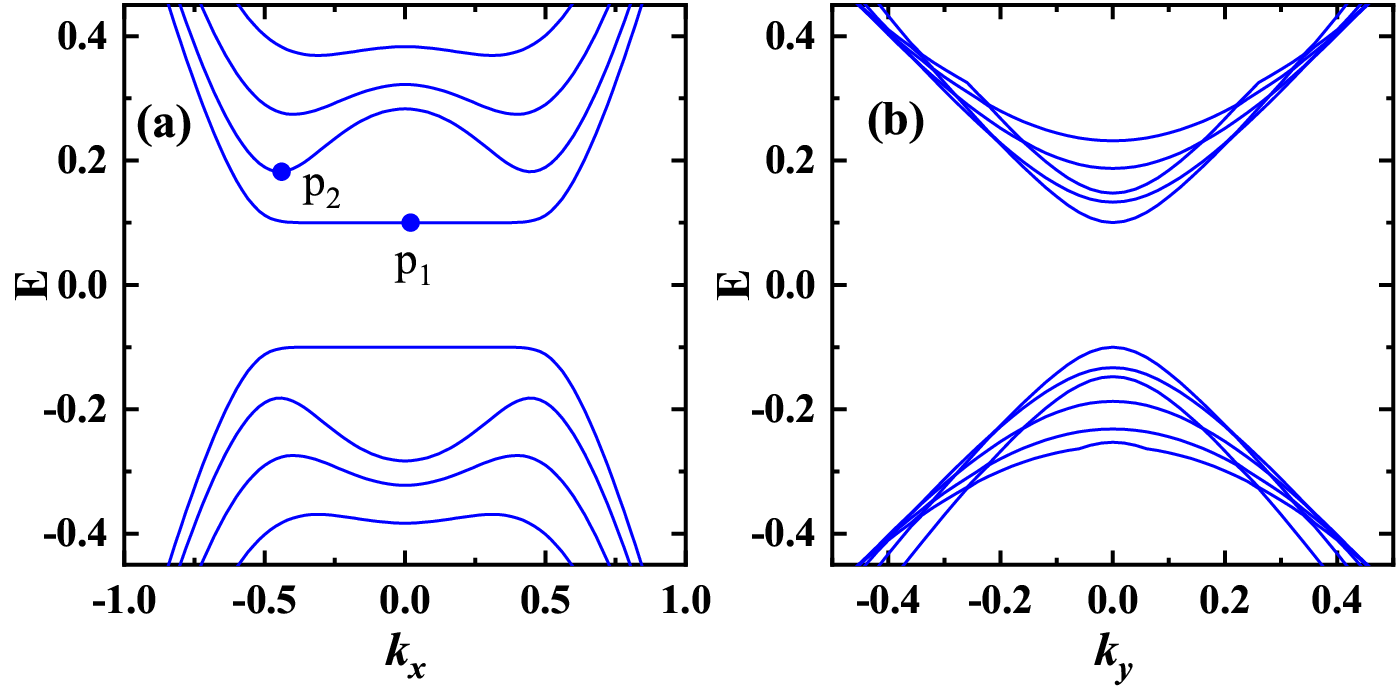}
\caption{Band structures of the tight-binding Hamiltonian in Eq.~(\ref{Ham2}) along the $k_x$ direction in (a) and along the $k_y$ direction in (b). $p_1$ and $p_2$ label the first and second conduction bands in $k_x$, respectively.}\label{fig2}
\end{figure}

In this section, we present numerical results for the linear thermal noise in the four-terminal system shown in Fig.~\ref{fig1}. We consider two typical setups. In setup {\color{blue}I} [panel (a)], the bias voltages applied to each terminal are described by the voltage profile ${\cal{V}}=(0,0,V/2,-V/2)$, while in setup {\color{blue}II} [panel (b)] the voltage profile is ${\cal{V}}=(V/2,-V/2,0,0)$. Correspondingly, the driving electric field is along $E_y$ in setup {\color{blue}I} and along $E_x$ in setup {\color{blue}II}. Without loss of generality, we set the bias amplitude to $V=0.2$ in the calculations. The central scattering region is a square lattice of size $N_x \times N_y = 30 \times 30$, which is described by the tight-binding Hamiltonian in Eq.~(\ref{Ham2}). This four-terminal system preserves $\cal{T}$ symmetry and mirror symmetry ${\cal{M}}_x$, which allows a BCD pseudovector $\mathcal{D}_x$ in the $x$ direction. The corresponding band structures are depicted in Fig.~\ref{fig2}. The dispersions along $k_x$ and $k_y$ are clearly different, due to the symmetry of the underlying system.


To compare with the linear thermal noises in bulk systems [Eqs.~(\ref{Sxx})--(\ref{Sxy})], we focus on auto-correlation functions $S^{(1)}_{11}$ and $S^{(1)}_{33}$, and cross-correlation $S^{(1)}_{13}$ and $S^{(1)}_{31}$. According to Eqs.~(\ref{Snoise4}), the voltage profile ${\cal{V}}=(0,0,V/2,-V/2)$ leads to $S^{(1)}_{12}=0$ in setup {\color{blue}I}, since linear noise $S^{(1)}_{12}$ depends on $V_1$ and $V_2$ and vanishes in this configuration. Similarly, in setup {\color{blue}II}, one has $S^{(1)}_{34}=0$. From the current conservation condition $\sum_{\gamma}S^{(1)}_{\alpha\gamma}=0$, we have
\begin{eqnarray}
&S^{(1)}_{11}& + S^{(1)}_{13} + S^{(1)}_{14} = 0  ~~\text{for setup {\color{blue}I}}, \label{CC1} \\
&S^{(1)}_{33}& + S^{(1)}_{31} + S^{(1)}_{32} = 0  ~~\text{for setup {\color{blue}II}}. \label{CC2}
\end{eqnarray}

\begin{figure}
\centering
\includegraphics[width=0.9\columnwidth]{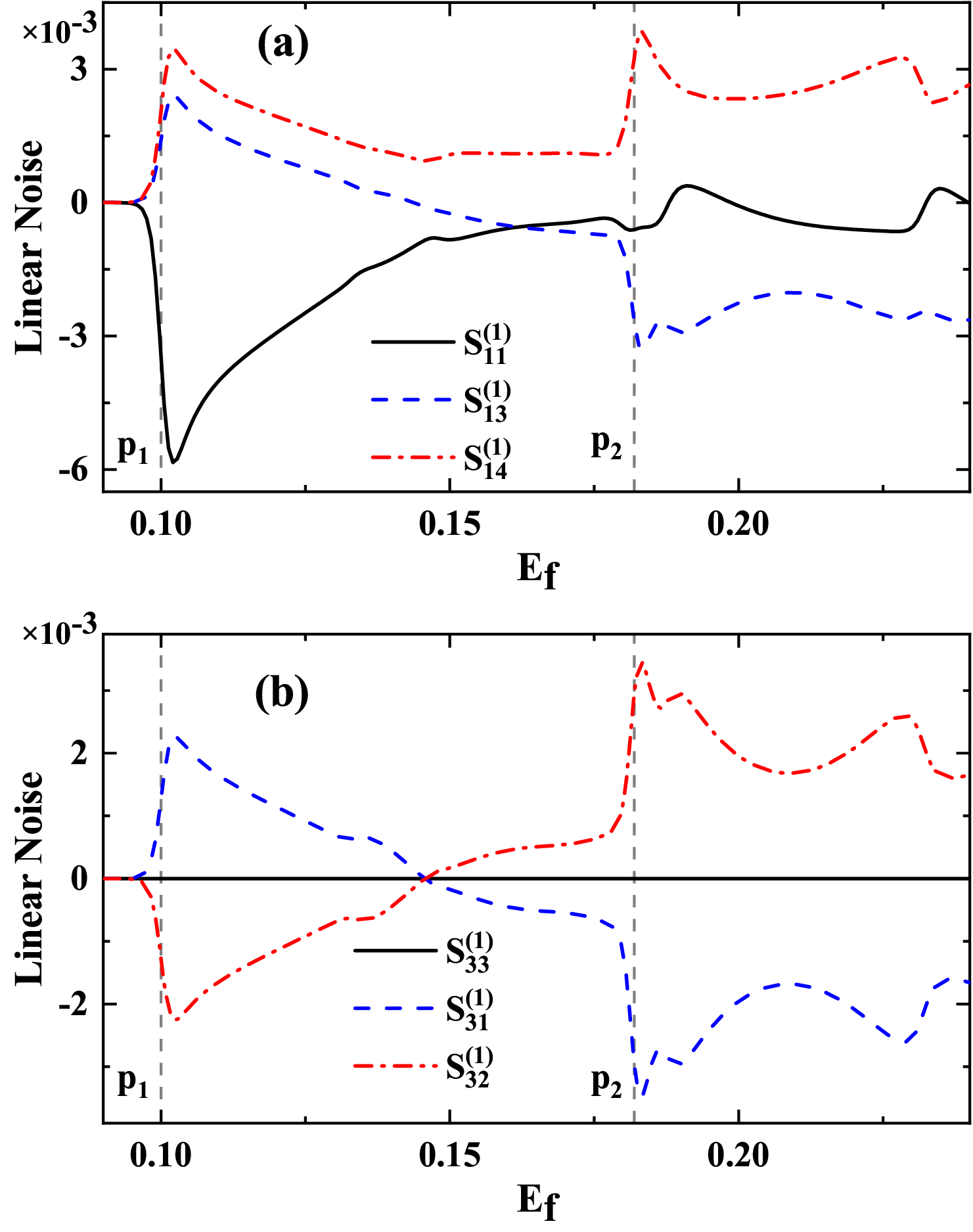}
\caption{Linear thermal noises as a function of the Fermi energy at $T=10$ $\text{K}$. Panel (a) corresponds to setup {\color{blue}I} with a bias in $x$ direction, while panel (b) is for setup {\color{blue}II} with a driving electric field in $y$ direction. $p_1$ and $p_2$ label the first and second band edges along $k_x$. }\label{fig3}
\end{figure}

Numerical results shown in Fig.~\ref{fig3} are in agreement with the above analysis. Fig.~\ref{fig3}(a) presents the linear thermal noise for setup {\color{blue}I} as a function of the Fermi energy $E_f$. Over the entire energy range considered, we numerically confirm that $S^{(1)}_{11} + S^{(1)}_{13} + S^{(1)}_{14} = 0$.
Now we focus on the auto-correlation $S^{(1)}_{11}$. By defining $M_{\alpha\beta} = \Gamma_\alpha G_0^r (u_\beta - I) G_0^r \Gamma_\beta G_0^a +\Gamma_\alpha G_0^r\Gamma_\beta G_0^a(u_\beta - I) G_0^a$ with $I$ the identity matrix, $S^{(1)}_{11}$ can be rewritten as
\begin{eqnarray}
S^{(1)}_{11} &=& -2 \int_{E}[ M_{13}V_3 + M_{14}V_4 ]g_2, \nonumber \\
             &=& 2 k_B T \int_{E} [ M_{13}V_3 + M_{14}V_4 ] \partial_E f_0, \label{S11}
\end{eqnarray}
where we have used $g_2 = -k_B T \partial_E f_0$. This expression shows that $S^{(1)}_{11}$ scales with $2 k_B T$, consistent with the linear noise ${\cal S}^{(1)}_{xx}$ for bulk systems in Eq.~(\ref{Sxx}). The energy dependence in Fig.~\ref{fig3}(a) reveals that $S^{(1)}_{11}$ is maximized near the band edge $p_1$ and then decreases as $E_f$ moves into the band. This behavior implies BCD-induced physics, since BCD is optimal at the band edge,~\cite{SodemannFu2015} which is also in agreement with the bulk noise ${\cal S}^{(1)}_{xx}$. The cross-correlations $S^{(1)}_{13}$ and $S^{(1)}_{14}$ exhibit similar band-edge enhancement near $p_1$ and $p_2$, consistent with BCD-induced mechanism. Although these cross-correlations have no direct bulk counterparts and are beyond the scope of semiclassical bulk theory,~\cite{Wei2023} the following expressions that show that both $S^{(1)}_{13}$ and $S^{(1)}_{14}$ scale with $k_B T$
\begin{eqnarray}
S^{(1)}_{13}= -k_B T \int_{E}M_{13}V_3 \partial_E f_0, \label{S13} \\
S^{(1)}_{14}= -k_B T \int_{E}M_{14}V_4 \partial_E f_0. \label{S14}
\end{eqnarray}
In the low-energy regime, $S^{(1)}_{13}$ and $S^{(1)}_{14}$ are positive and decrease with increasing $E_f$. As $E_f$ approaches $p_2$, $S^{(1)}_{13}$ changes sign and becomes nearly antisymmetric to $S^{(1)}_{14}$. In addition, $S^{(1)}_{13}$ is not equal to $S^{(1)}_{14}$, which arises from symmetry breaking of the four-terminal system in the $y$ direction.


Figure~\ref{fig3}(b) shows the linear thermal noise for setup {\color{blue}II} with a driving field $E_x$. We find that $S^{(1)}_{33} + S^{(1)}_{31} + S^{(1)}_{32} = 0$, which satisfies current conservation. The cross-correlations $S^{(1)}_{31}$ and $S^{(1)}_{32}$ peak near the band edges $p_1$ and $p_2$, which correspond to BCD-induced behavior. Notice that $S^{(1)}_{33}=0$. To understand this, we rewrite its expression as
\begin{align}
S^{(1)}_{33} = 2 k_B T \int_{E}[ M_{31}V_2 + M_{32}V_1 ] \partial_E f_0.
\end{align}
Since the system has mirror symmetry ${\cal M}_x$, exchanging $V_1$ and $V_2$ should give the same result, which shows
\begin{align}
S^{(1)}_{33} = 2 k_B T \int_{E}[ M_{31}V_1 + M_{32}V_2 ] \partial_E f_0. \label{S33}
\end{align}
Then we have $M_{31}=M_{32}$. In setup {\color{blue}II}, $V_1$ = $-V_2$ and hence $S^{(1)}_{33}=0$. This vanishing behavior $S^{(1)}_{33}=0$ is the multi-terminal counterpart of ${\cal S}^{(1)}_{yy} = 0$ for bulk systems in Eq.~(\ref{Syy}). In addition, the mirror symmetry ${\cal M}_x$ together with $V_1$ = $-V_2$ enforces $S^{(1)}_{31} = -S^{(1)}_{32}$, where
\begin{align}
S^{(1)}_{31}= -k_B T \int_{E}M_{31}V_1 \partial_E f_0, \label{S31} \\
S^{(1)}_{32}= -k_B T \int_{E}M_{32}V_2 \partial_E f_0, \label{S32}
\end{align}
Notice that both $S^{(1)}_{31}$ and $S^{(1)}_{32}$ correspond to the bulk cross-correlation ${\cal S}^{(1)}_{xy}$ in Eq.~(\ref{Sxy}), and scale as $k_B T$.

\begin{figure}
\centering
\includegraphics[width=0.8\columnwidth]{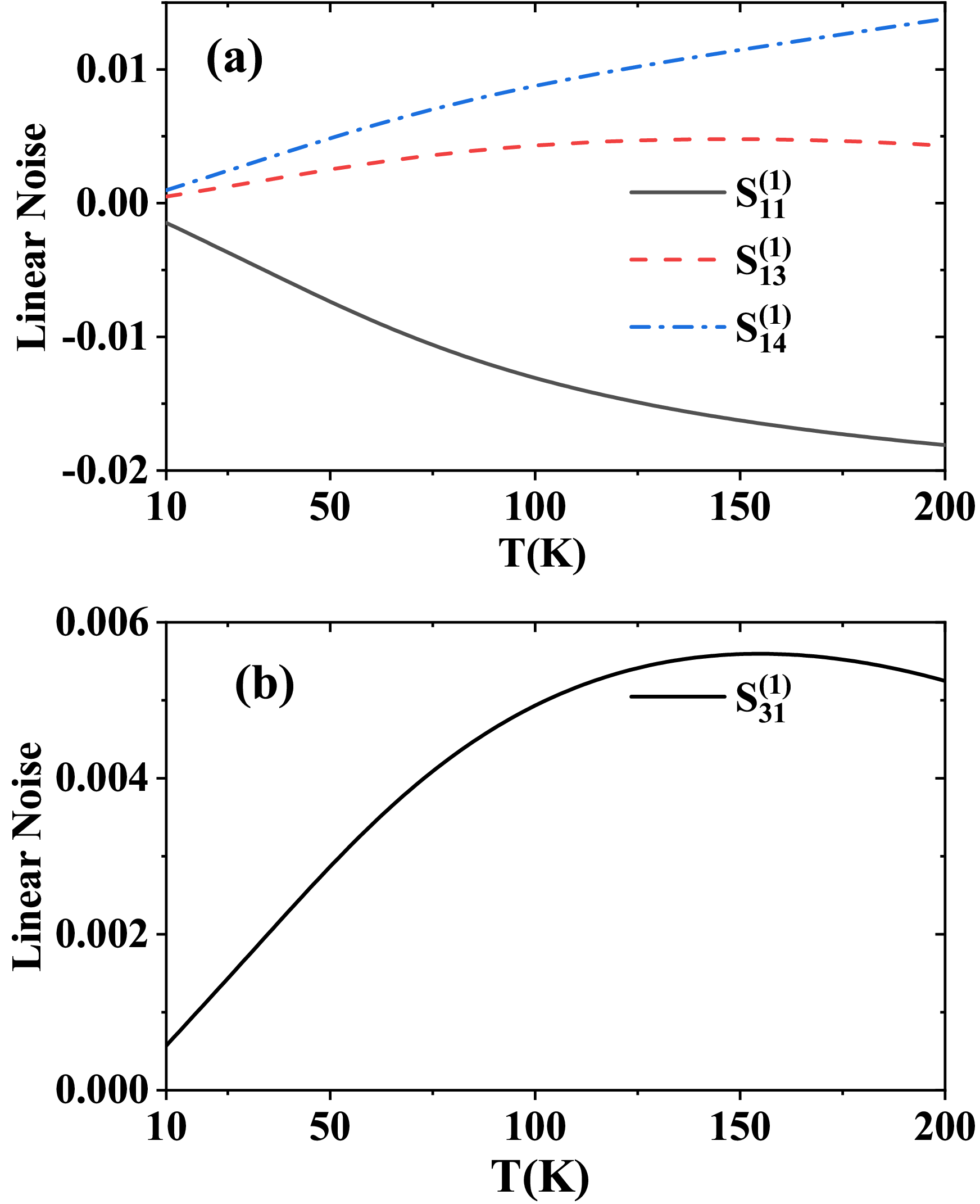}
\caption{Linear thermal noises as a function of the temperature at $E_f=0.12$. (a) $S^{(1)}_{11}$, $S^{(1)}_{13}$ and $S^{(1)}_{14}$ for setup {\color{blue}I}. (b) $S^{(1)}_{31}$ for setup {\color{blue}II}. }\label{fig4}
\end{figure}

Remarkably, the finite $S^{(1)}_{11}$ in Eq.~(\ref{S11}), vanishing $S^{(1)}_{33}$ in Eq.~(\ref{S33}), and cross-correlations $S^{(1)}_{31}$ and $S^{(1)}_{32}$ in Eqs.~(\ref{S31})--(\ref{S32}) together reveal the symmetry(geometry)-selection rule of linear thermal noise. The auto-correlation scales as $2 k_B T$ when the driving field $E_y$ is perpendicular to the BCD vector ${\cal D}_x$ and vanishes when the field $E_x$ is parallel to ${\cal D}_x$, while the cross-correlations scale as $k_B T$. All of these signals show pronounced peaks near the band edges, which are the signature of BCD-induced physics.~\cite{SodemannFu2015} These observations show one-to-one correspondence with the semiclassical results in Eqs.~(\ref{Sxx})-(\ref{Syy}). In previous work, we established the one-to-one correspondence between quantum multi-terminal theory and semiclassical theory on the BCD-induced second-order nonlinear Hall effect.~\cite{Wei2022} The present study extends this correspondence to BCD-induced linear thermal noise. Moreover, the quantum multi-terminal theory provides more information than the semiclassical bulk theory. In the semiclassical results [Eqs.~(\ref{Sxx})--(\ref{Sxy})]~\cite{Wei2023}, the cross-correlation is defined between current densities in two spatial directions, whereas in the quantum multi-terminal theory it refers to the correlation between currents in different terminals. Importantly, the terminal-resolved correlations are directly measurable, enabling direct comparison with experiments.~\cite{Djukic2006,Tamir2022}

We further investigate the temperature dependence of the linear thermal noise. For a fixed Fermi energy $E_f=0.12$, we calculate the linear thermal noises in setup {\color{blue}I} and setup {\color{blue}II} over a range of temperatures and present the numerical results in Fig.~\ref{fig4}. At low temperatures ($T<50$ $\text{K}$), the noises scale almost linearly with $T$, which is consistent with the semiclassical bulk theory [Eqs.~(\ref{Sxx})--(\ref{Sxy})].~\cite{Wei2023} As $T$ increases, the curves gradually flatten. In particular, for $T > 150$ $\text{K}$, $S^{(1)}_{31}$ in setup {\color{blue}II} begins to decrease. The deviation from linear scaling is consequence of thermal broadening: as the Fermi distribution function widens with increasing $T$, the energy integrals in Eqs.~(\ref{Snoise4}) is performed in a larger energy interval and no longer increase linearly with temperature.

\begin{figure}[tbp]
\centering
\includegraphics[width=0.8\columnwidth]{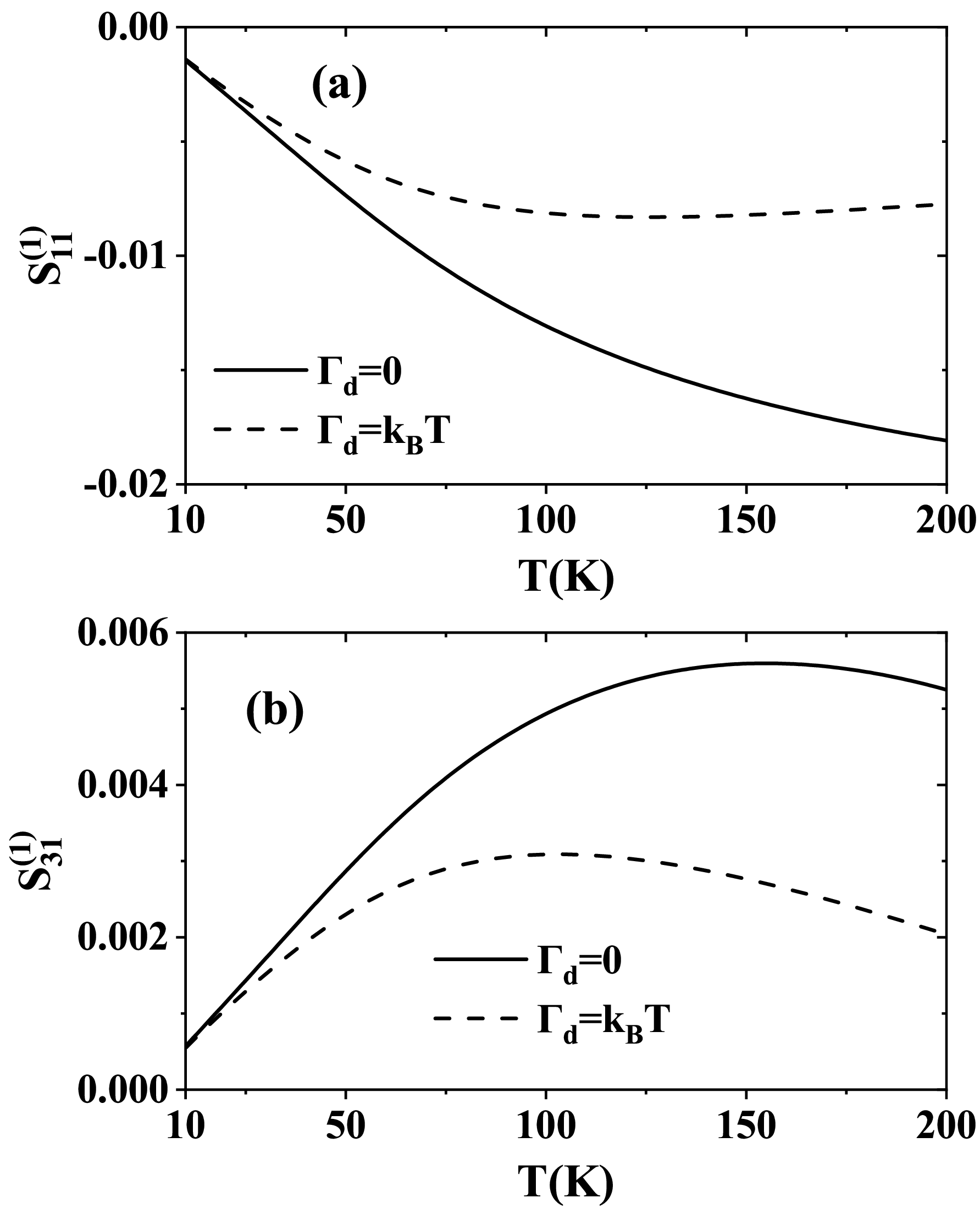}
\caption{Dephasing effect on the linear thermal noise. (a) $S^{(1)}_{11}$ for setup {\color{blue}I}. (b) $S^{(1)}_{31}$ for setup {\color{blue}II}.}\label{fig5}
\end{figure}

At finite temperature, lattice vibrations are thermally excited as phonons. Electro--phonon coupling provides a source of inelastic scattering and acts as a dephasing mechanism that hampers coherent transport. Although the linear thermal noise generally increases with temperature, its transport behavior can be modified by phonon-induced dephasing. Therefore, it is necessary to evaluate the impact of dephasing on thermal noise at finite temperature. We adopt the imaginary potential model~\cite{Datta1991,Efetov1995,Brouwer1997} to simulate the dephasing effect. In this model, an on-site imaginary potential $-i\Gamma_d/2$ is introduced in the central scattering region, where $\Gamma_d = k_B T$ characterizes the dephasing strength at temperature $T$. The numerical results are shown in Fig.~\ref{fig5}. We find that the auto-correlation $S^{(1)}_{11}$ decreases as the dephasing strengthens with increasing temperature, and it drops by about half at $T = 100~K$ [Fig.~\ref{fig5}(a)]. The cross-correlation $S^{(1)}_{31}$ declines at high temperature in the absence of dephasing, while the dephasing effect lowers the onset for the decline of $S^{(1)}_{31}$ [Fig.~\ref{fig5}(b)]. Therefore, to observe a pronounced noise signal, the optimal temperature regime is $T<50$ $\text{K}$, where linear thermal noise increases almost linearly with $T$ and the dephasing effect remains weak.

\section{Conclusion}


In summary, we develop a quantum multi-terminal theory satisfying current conservation and gauge invariance, and investigate linear thermal noise in a four-terminal Hall setup with a finite Berry curvature dipole (BCD). By comparing with the semiclassical results for bulk systems, we demonstrate the one-to-one correspondence between terminal-resolved linear noise $S^{(1)}_{\alpha\beta}$ and direction-resolved bulk noise ${\cal S}^{(1)}_{ab}$. Both the auto-correlation functions $S^{(1)}_{11}$ and ${\cal S}^{(1)}_{xx}$ scale as $2 k_B T$ when the driving field $E_y$ is perpendicular to the BCD vector ${\cal D}_x$, and vanish when $E_x \parallel {\cal D}_x$. In contrast, when $E_x \parallel {\cal D}_x$, the cross-correlations $S^{(1)}_{31}$ and ${\cal S}^{(1)}_{xy}$ scale as $ k_B T$. Upon varying the Fermi energy, the terminal-resolved linear noise shows peaks near the band edges where the BCD is optimal, consistent with the semiclassical prediction for bulk systems. Finally, we find that dephasing suppresses the noise at elevated temperatures, indicating an optimal observation window in the low-temperature regime ($T < 50$ K), where the BCD-induced signal remains pronounced. Our work establishes a direct connection between semiclassical bulk theory and quantum multi-terminal theory of linear thermal noise, and reveals the role of symmetry in constraining its transport properties.


\section*{acknowledgments}

This work was supported by the National Natural Science Foundation of China (Grants No.~12574054, No.~12404058, and No.~12034014). F.X. also acknowledges the
Shenzhen Science and Technology Program (Grant No. JCYJ20220818100204010).


\end{document}